# Band-Gap-Dependent Electronic Compressibility of Carbon Nanotubes in the Wigner Crystal Regime

Neda Lotfizadeh[1], Daniel R. McCulley[2], Mitchell J. Senger[2], Han Fu[3], Ethan D. Minot[2], Brian Skinner[4,5] and Vikram V. Deshpande[1]*

[1]Department of Physics and Astronomy, University of Utah, Salt Lake City, Utah 84112, USA.

[2]Department of Physics, Oregon State University, Corvallis, OR 97331, USA.

[3]James Franck Institute, University of Chicago, Chicago, Illinois 60637, USA.

[4]Department of Physics, Massachusetts Institute of Technology, Cambridge, MA 02139, USA.

[5]Department of Physics, Ohio State University, Columbus, Ohio 43210, USA.

*Correspondence to: vdesh@physics.utah.edu.

**Electronic compressibility, the second derivative of ground state energy with respect to total electron number, is a measurable quantity that reveals the interaction strength of a system and can be used to characterize the orderly crystalline lattice of electrons known as the Wigner crystal. Here, we measure the electronic compressibility of individual suspended ultraclean carbon nanotubes in the low-density Wigner crystal regime. Using low-temperature quantum transport measurements, we determine the compressibility as a function of carrier number in nanotubes with varying band gaps. We observe two qualitatively different trends in compressibility versus carrier number, both of which can be explained using a theoretical model of a Wigner crystal that accounts for both the band gap and the confining potential experienced by charge carriers. We extract the interaction strength as a function of carrier number for individual nanotubes and show that the compressibility can be used to distinguish between strongly and weakly interacting regimes.**

The Wigner crystal, an ordered crystalline lattice of electrons with extremely strong interactions, is one of the most fascinating regimes of solid-state physics [1]. One of the observables of this regime is the electronic compressibility $\kappa$, which is a reflection of the many-body interactions of the target system and can be obtained from $\kappa = (d^2E/dN^2)^{-1} = (d\mu/dN)^{-1}$, where $E$ is the ground-state energy, $N$ is the total electron number, and $\mu = dE/dN$ is the chemical potential of the system. The inverse compressibility $\kappa^{-1}$, in particular, corresponds to the amount by which the chemical potential must be raised in order to add an electron; small $\kappa^{-1}$ indicates that the system easily accommodates additional electrons. Various studies have been conducted on compressibility (and quantum capacitance, which is directly proportional to compressibility) of quantum structures such as quantum wires, two-dimensional electron systems, mono- and bilayer graphene, etc. to explore the interactions in these systems [2–6]. When the density of states is constant and electron-electron (*e-e*) interactions are relatively weak, the compressibility of an electronic system is independent of the charge carrier density. These assumptions are violated, however, in the low-density regime, and the compressibility varies strongly with density. In particular, in the Wigner crystal regime strong correlations between electrons are predicted to lead to a sharp decrease in $\kappa^{-1}$ with decreasing density [2,7,8]. Previous studies have indeed observed a reduction in $\kappa^{-1}$ at low densities in macroscopic (i.e. laterally unconfined) structures [9–13]; this trend has been attributed to strong screening effects from a nearby metal gate, the presence of disorder in the system, or contributions from the exchange interaction. Unfortunately, for meso- or nano-scale systems the downward trend in $\kappa^{-1}$ is easily reversed by the effect of an electrostatic confining potential produced by gate and source/drain electrodes, which tends to push electrons into an even smaller spatial region as their density is reduced. To our best knowledge, suppression in $\kappa^{-1}$ at low densities has never been reported in laterally confined quantum structures.

Suspended carbon nanotubes (CNTs) are a promising platform for investigating the effects of strong electronic correlations in one dimension. As a clean, interacting quantum system, electrons in a suspended CNT at low density [14] may be described as a Wigner crystal [15]. Indeed, experimental studies have confirmed fascinating magnetic and electronic properties of the Wigner crystal phase, such as their exponentially suppressed exchange energy [16], absence of excited energy states [17] and giant orbital magnetic moment [18]. These observations indicate that despite more than a decade of studies on the 1D Wigner crystal, improvement in device fabrication and higher quality carbon nanotubes lead to the discovery of novel signatures that have not been revealed before. Very recently, Shapir *et al.* [19] have developed a technique to observe the Wigner crystal directly by imaging the charge density of the system in real space. Providing detailed theoretical calculations, they showed that the Wigner crystal regime has one of the strongest *e-e* interactions in the solid state. The strength of interactions is usually parameterized by $r_s$, defined as the ratio of the Coulomb interaction between electrons separated by a distance *r*, $e^2/r$, to their typical kinetic energy $h^2/(m^*r^2)$. The interaction strength can be written in terms of the effective Bohr radius $a_B$ as $r_s \approx 1/(na_B)$, where *n* is the one-dimensional electron density. The effective mass *m\** is proportional to the CNT bandgap, $\Delta$, (with $\Delta = 2m^*v^2$, where *v* is the Fermi velocity), so that increasing $\Delta$ leads to a larger $r_s$, and a stronger role for interactions. Previous studies of the addition energy spectrum, capacitance and compressibility of CNTs have mostly focused on non-interacting physics and the weak-interaction regime [14,20–22]. But the effect of bandgap on these quantities in the Wigner crystal regime has not yet been considered.

In this work, we have studied one-dimensional systems with different *e-e* interaction strengths, using long suspended CNTs of various bandgaps. We report two contrasting trends of enhancement and suppression of $\kappa^{-1}$. In CNTs with very large bandgaps, we observe suppression of $\kappa^{-1}$ at low densities, and provide a theory to show how this trend can be produced by a Wigner crystal. Using this theory, we show that compressibility is sensitive to both bandgap and confining potential of the nanotube, which provides insight into the electronic interactions in these materials.

Our CNTs are grown using chemical vapor deposition across a ~2 μm wide trench on prefabricated substrates to eliminate disorder effects (see Supplemental Material [23]) [24]. A pair of gate electrodes is at the bottom of the trench and ~750 nm below the contact electrodes. Fig. 1a shows a schematic of the device.

We focus on the low-density regime of electrons or holes in clean CNTs, which clearly exhibits single-electron/hole conductance peaks in the Coulomb blockade (CB) regime, down to the last electron/hole at the conduction/valence band edge. The charge carrier density of CNTs can be modulated using electrostatic gating. A high-resolution map of the differential conductance *dI/dV* as a function of gate voltage $V_g$ and source-drain bias voltage $V_{sd}$ is shown in in Fig. 1b for T = 1.5 K, and illustrates CB diamonds and a bandgap of Δ ~ 25 meV in CN1. Fig. 1c plots the conductance of CN1 as a function of $V_g$. The regularity of CB peaks in this data, as well as the electron interference patterns in our devices with more transparent contacts [25] made with the same procedure, indicates that our devices are high-quality and defect-free. In Fig. 1c, the CB peaks get closer going from low to high carrier number. Figure 1d shows similar data from another device (CN2, with Δ ~ 165 meV); the CB peaks in CN2 show the opposite behavior, i.e. the CB peaks spread further apart with increasing carrier number.

The compressibility of the nanotube can be obtained from gate voltage spacing between the neighboring CB peaks in the transport data converted to energy: $\delta_N = E_{N+1} - 2E_N + E_{N-1} = \kappa^{-1}$ see e.g. [26,27], using $\mu = \alpha e V_g$, where gate voltage lever arm $\alpha = V_c/V_g$, and $V_c$ is the height of rhombic pattern in the $G(V_g, V_{sd})$ diagram [28]. Figures 1e and 1f plot the extracted value of $\kappa^{-1}$ as a function of carrier number for CN1 and CN2. The alternating pattern in some parts of the plots arises from filling the subsequent orbital states with two electrons having opposite spins [14,20]. In CN1 ($\Delta \sim 25$ meV), $\kappa^{-1}$ is higher at low densities. This trend of addition energy has been reported previously and explained using a single-particle picture [14,20,21]. Due to the small effective mass of CN1, the energetics in this device has been considered to be dominated by a classical charging energy and the quantum kinetic energy. It is worth noting that the device imaged by Shapir *et al.* [19] with $\Delta = 45$ meV has similar energetics to CN1 and was found to be a Wigner crystal. On the other hand, we observe the opposite trend in CN2 with $\Delta \sim 165$ meV; in this device $\kappa^{-1}$ is suppressed at low densities. In contrast to CN1, the effective mass of CN2 is large and the energetics are more likely to be dominated by Coulomb interactions. Correspondingly, the electronic compressibility of a Wigner crystal may follow a different trend in samples with such large gaps.

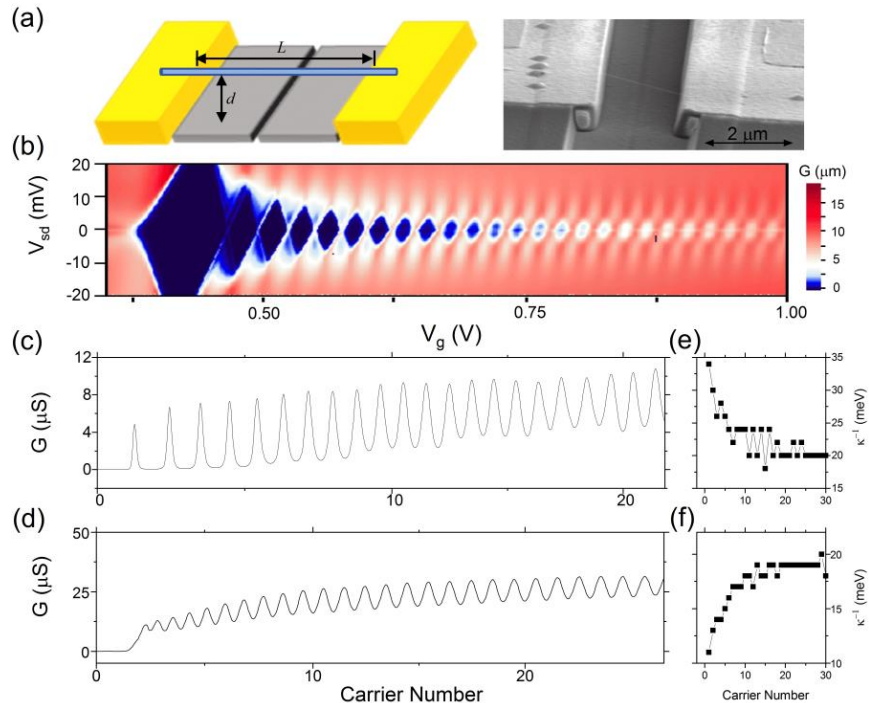

**Fig. 1:** (a) Schematic diagram (Left) and SEM image of a device (Right). Carbon nanotube is suspended over a 2 μm-wide trench. The vertical spacing between trench and contact electrodes is $d = 760$ nm. (b) Color scale plot of differential conductance versus gate voltage $V_g$ and source-drain bias $V_{sd}$ in CN1. Conductance $G$ versus carrier number in (c) CN1 and (d) CN2. (e,f) Inverse compressibility as a function of carrier number for the related device.

It is desirable to vary the bandgap parameter to study its effect on $\kappa^{-1}$. One way to do this in a continuous manner is by applying an external magnetic field ($B$) parallel to the axis of the tube [29–31]. This is particularly applicable to CN1 which has a small bandgap at $B = 0$ and can display field-dependent energetics. Figure 2 shows $\kappa^{-1}$ in CN1 as the magnetic field is varied from $B = 0.4$ T, to $B = 4$ T. The minimum bandgap, $\Delta_{min}$, is obtained at 0.4 T and at higher fields the gap increases at a rate of ~2.5 meV/T. As the bandgap is increased, $\kappa^{-1}$ is observed to decrease at lowest densities.

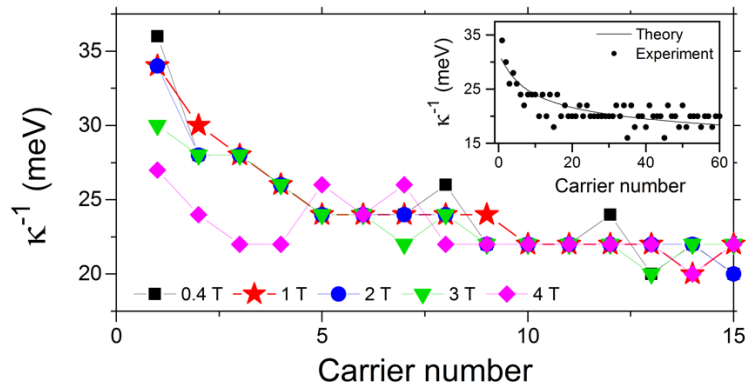

**Fig. 2:** Effect of magnetic field on $\kappa^{-1}$ of CN1 for a range of magnetic fields, from $B= 0.4$ T, where the bandgap reaches its minimum, to 4 T. Inset: Comparison of theoretical (solid line) and experimental (dots) results of $\kappa^{-1}$ as a function of carrier number for CN1.

To study the suppression in $\kappa^{-1}$ at low densities, similar to CN2, we examine a range of different samples with appropriately large bandgaps. Fig. 3a shows the measured $\kappa^{-1}$ as a function of carrier number in five devices (CN2-CN6) with bandgaps $\geq$ 150 meV. In all these samples, we observe the same trend as in CN2, meaning that in these tubes $\kappa^{-1}$ is suppressed by going to low densities.

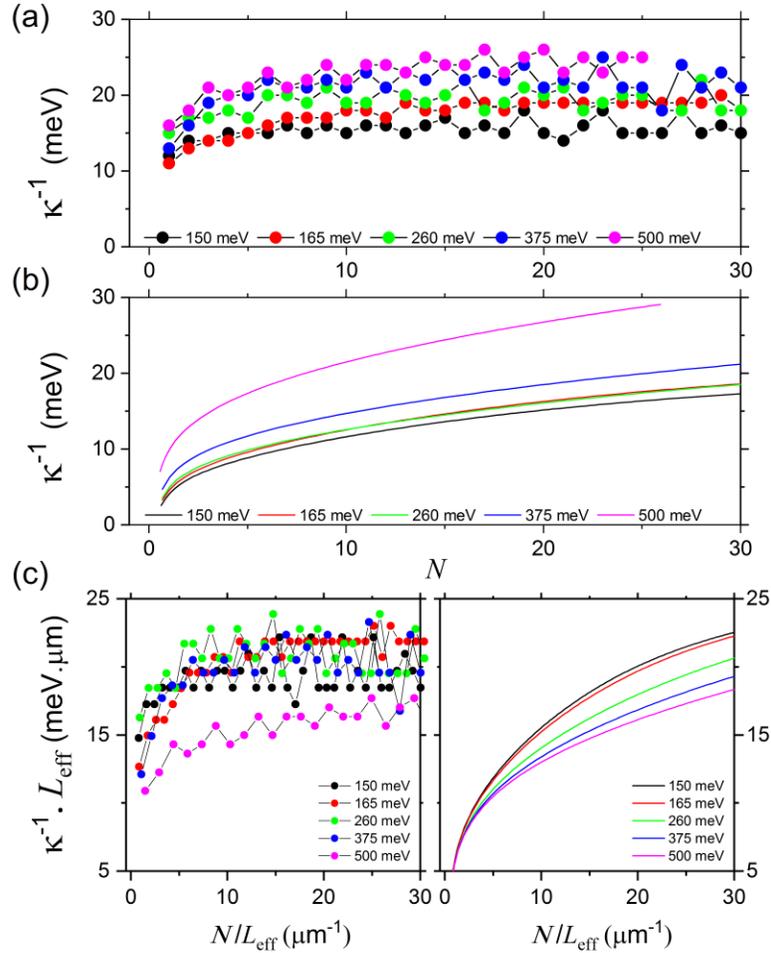

**Fig. 3.** Experimental data (a) and theoretical results (b) of $\kappa^{-1}$ as a function of carrier number ($N$) for nanotubes with different bandgaps. (c) Experimental (Left) and theoretical (Right) results converted to $\kappa^{-1}$ times $L_{\text{eff}}$ as a function of charge density.

To understand behaviors of $\kappa^{-1}$ at low densities, we propose an interacting model in which we calculate the ground state energy $E$ of a system having $N$ electrons using the Hamiltonian $H = \sum_i K_i + \sum_{i \neq j} V(r_{ij}) + \sum_i U(r_i)$, where $K_i$ is the kinetic energy operator for electron $i$, $V(r_{ij})$ is the interaction energy between two electrons separated by a distance $r_{ij}$, and $U(x)$ is the potential energy of an electron at position $x$ due to an external electric potential. In the case of $U(x) \equiv 0$ electrons are arranged with a uniform (voltage-dependent) density $n$ along a line of length $L_{\text{eff}}$, so that the total number of electrons in the system is $N = nL_{\text{eff}}$. In the Wigner crystal limit, the electrostatic energy $E_{\text{el}}$ of the system can be approximated by that of a classical collection of point charges with regular spacing $1/n$. In the limit where $L_{\text{eff}}$ is much longer than the distance $d$ to the gate electrode, $E_{el} = N \sum_{i=1}^{\infty} V(i/n)$, where the interaction energy $V(r)$ is given by the gate-screened Coulomb repulsion $V(r) = \frac{e^2}{4\pi\varepsilon_0}\left(\frac{1}{r} - \frac{1}{\sqrt{r^2+(2d)^2}}\right)$. At low electron densities $nd \ll 1$, the typical interaction energy becomes that of a dipole-dipole interaction, $V\left(\frac{1}{n}\right) \sim e^2 n^3 d^2/(4\pi\varepsilon_0)$. This rapid vanishing of $V$ with $n$ at low density implies that the electrostatic cost of inserting an additional energy decreases with decreasing concentration in cases where the electron density is uniform.

In the limit where *e-e* interactions dominate over the quantum kinetic energy of electrons and system adopts a Wigner crystal-like arrangement, the kinetic energy can be treated as a perturbation. In this situation, electron wave functions have little spatial overlap with each other and one can approximate the kinetic energy via a description where each electron is confined into a box of width $1/n$, such that neighboring electrons have no wave function overlap. Total kinetic energy is therefore given by the number of electrons multiplied by ground state energy of a 1D particle in a box. We describe the kinetic energy of an electron via the relativistic dispersion relation $\varepsilon(p) = \sqrt{(vp)^2 + (\Delta/2)^2} - \Delta/2$, where $v$ is the Fermi velocity, $p$ is the electron momentum and the bandgap is $\Delta = 2m^*v^2$. Note that at low electron densities with small $p$, $\varepsilon(p)$ reduces to the familiar form of $\varepsilon(p) \cong p^2/(2m)$. Our approximation of a Coulomb-dominated electron state is justified when $V(1/n)$ is much larger than the typical kinetic energy scale $\varepsilon(p = \pi\hbar n)$. At low electron densities this inequality is satisfied in the usual limit of large $r_s$, $na_B \ll 1$. In our experiments $a_B$ is no larger than ~ 15 nm, while our CNT lengths are of order 2 μm, so our approximation is justified when there are fewer than ~ 100 electrons in the system. The inverse compressibility is then: $\kappa^{-1} = (1/L_{eff})d\mu/dn$. A detailed expression for $\mu$ is given in the Supplemental Material. Our theoretical results for $\kappa^{-1}$, illustrated in Fig. 3b, have the same trend and magnitude as our transport data for CN2-CN6. These modeling results demonstrate that our theory achieves an expected explanation for the behavior of $\kappa^{-1}$ in the large bandgap devices.

In addition to the Coulomb interactions between electrons, the electric potential difference between the gate and source/drain electrodes creates an external potential that may significantly affect the compressibility of the system. To model this effect, we assume that electrons reside in the minimum of a potential well described generically by $U(x) = \frac{e^2}{4\pi\varepsilon_0}(\frac{x^2}{D_1^3} + \frac{x^4}{D_2^5})$. The position $x$ is defined with respect to the location of the potential minimum, and the value of the electrostatic potential at this minimum can be set to zero. The length scales $D_1$ and $D_2$ define the strength of the potential. In the presence of such a confining potential the electron density varies with position $x$, with electrons being more densely spaced at $x = 0$ and more sparsely spaced at larger distances from the minimum of confining potential [7].

Fig. 2 inset compares the result of our theoretical calculation (solid line) and the measured data (dots) for CN1. Our fitted parameters $D_1$ and $D_2$ agree with the estimated value from electrostatic calculations in ref [7]. It can be seen that the theory matches very well with the experiment, implying that the confining potential plays an important role in the enhancement of $\kappa^{-1}$ at low density for this tube. The larger values of $\kappa^{-1}$ with decreasing $N$ suggest that electrons are pushed together by the confining potential, so that $L_{\text{eff}}$ of the device increases with increasing $N$. We have also calculated $\kappa^{-1}$ in the presence of magnetic field by adding a field-tunable gap [27] $\Delta_B$ to the non-vanishing gap $\Delta_{min}$, and in the presence of a confining potential, we are able to derive a change in $\kappa^{-1}$ as a function of $B$ that is qualitatively similar to our experimental results (see Fig. S2). The opposite trend of $\kappa^{-1}$ observed in the larger bandgap devices (CN2-CN6) suggests that the confining potential plays a weaker role in those devices, compared to CN1.

Obtained values of $L_{\text{eff}}$ and $a_B$ for individual tubes from our theory are presented in TableS1. The values of $r_s$ range from ~ 3 (for small bandgap and large $N$) to ~ 450 (for large bandgap and small $N$). Large $r_s$ is consistent with our initial assumption of the Wigner crystal regime [19] and justify our estimate that contact interactions are negligible [33] in our large bandgap tubes.

Interactions in the low-density regime are stronger in tubes with larger $\Delta \times L_{\text{eff}}$. In order to eliminate the effect of $L_{\text{eff}}$ and present compressibility dependence on bandgap of CNTs, results of Fig. 3a,b are illustrated in Fig. 3c in terms of $\kappa^{-1} \cdot L_{\text{eff}}$ as a function of density. Levitov and Tsvelik [8] had previously theorized that large bandgap tubes with slowly increasing $\kappa^{-1}$ (with increasing density) are more strongly interacting. According to our measurements and calculations in Fig. 3c, our tubes with larger bandgap reach the constant $\kappa^{-1}$ regime slower than CNTs with smaller bandgap, which is consistent with ref [8]. Overall, our devices show the same behavior as our model, indicating that *e-e* interactions are stronger in low density regime of nanotubes with larger bandgaps, causing $\kappa^{-1}$ to grow with density.

The observed compressibility behavior by itself is not proof of a Wigner crystal. Previous works had explained a similar suppression of compressibility as a function of density, though not in a laterally confined structure, based on the exchange interaction in a uniform gas ($r_s = 0$) model [10,12]. The observed behavior in our devices could also be described using the simple model of a uniform electron gas with exchange interaction (presented in the Supplemental Material). However, given the overwhelming evidence for Wigner crystallization from other experiments [16–19] in the parameter space of our devices, we can safely suggest our observed compressibility behavior as a probe of interaction strength of 1D Wigner crystals. Future studies will incorporate independent control of bandgap and confining potential.

In summary, we studied the effect of interactions on electronic compressibility of carbon nanotubes with different bandgaps. We showed that contact interactions are not negligible in tubes with smaller bandgaps and their compressibility can be tuned by applying external magnetic field. For stronger (weaker) interactions, inverse compressibility decreases (increases) in the limit of low density in the Wigner crystal regime. In devices with addition energy suppression at low density, tubes with larger bandgaps reach the noninteracting regime at larger densities compare to tubes with smaller bandgaps. Our theoretical modeling suggests that we are in a regime of relatively large $r_s$, and our data is consistent with a theoretical model of a Wigner crystal in a soft confining potential.

**Acknowledgments**

B.S. was supported by the NSF STC "Center for Integrated Quantum Materials" under Cooperative Agreement No. DMR-1231319. Work performed in Oregon was supported by the National Science Foundation under Grant No. 1709800. A portion of device fabrication was carried out in the University of California Santa Barbara (UCSB) nanofabrication facility.

## Competing interests

The authors declare no competing interests.

## References

[1] E. Wigner, Phys. Rev. **46**, 1002 (1934).
[2] M. M. Fogler, Phys. Rev. B **71**, 161304(R) (2005).
[3] L. A. Ponomarenko, R. Yang, R. V. Gorbachev, P. Blake, A. S. Mayorov, K. S. Novoselov, M. I. Katsnelson, and A. K. Geim, Phys. Rev. Lett. **105**, 136801 (2010).
[4] E. A. Henriksen and J. P. Eisenstein, Phys. Rev. B **82**, 041412(R) (2010).
[5] A. F. Young, C. R. Dean, I. Meric, S. Sorgenfrei, H. Ren, K. Watanabe, T. Taniguchi, J. Hone, K. L. Shepard, and P. Kim, Phys. Rev. B **85**, 235458 (2012).
[6] K. Steffen, R. Frésard, and T. Kopp, Phys. Rev. B **95**, 035143 (2017).
[7] H. Fu, B. I. Shklovskii, and B. Skinner, Phys. Rev. B **91**, 155118 (2015).
[8] L. S. Levitov and A. M. Tsvelik, Phys. Rev. Lett. **90**, 016401 (2003).
[9] J. P. Eisenstein, L. N. Pfeiffer, and K. W. West, Phys. Rev. Lett. **68**, 674 (1992).
[10] J. P. Eisenstein, L. N. Pfeiffer, and K. W. West, Phys. Rev. B **50**, 1760 (1994).
[11] L. Li, C. Richter, S. Paetel, T. Kopp, J. Mannhart, and R. C. Ashoori, Science **332**, 825 (2011).
[12] S. Shapira, U. Sivan, P. M. Solomon, E. Buchstab, M. Tischler, and G. Ben Yoseph, Phys. Rev. Lett. **77**, 3181 (1996).
[13] J. M. Riley, W. Meevasana, L. Bawden, M. Asakawa, T. Takayama, T. Eknapakul, T. K. Kim, M. Hoesch, S.-K. Mo, H. Takagi, T. Sasagawa, M. S. Bahramy, and P. D. C. King, Nature Nanotech **10**, 1043 (2015).
[14] S. Sapmaz, P. Jarillo-Herrero, L. P. Kouwenhoven, and H. S. J. van der Zant, Semicond. Sci. Technol. **21**, S52 (2006).
[15] V. V. Deshpande, M. Bockrath, L. I. Glazman, and A. Yacoby, Nature **464**, 209 (2010).
[16] V. V. Deshpande and M. Bockrath, Nature Phys **4**, 314 (2008).
[17] S. Pecker, F. Kuemmeth, A. Secchi, M. Rontani, D. C. Ralph, P. L. McEuen, and S. Ilani, Nature Phys **9**, 576 (2013).
[18] J. O. Island, M. Ostermann, L. Aspitarte, E. D. Minot, D. Varsano, E. Molinari, M. Rontani, and G. A. Steele, Phys. Rev. Lett. **121**, 127704 (2018).
[19] I. Shapir, A. Hamo, S. Pecker, C. P. Moca, Ö. Legeza, G. Zarand, and S. Ilani, Science **364**, 870 (2019).
[20] P. Jarillo-Herrero, S. Sapmaz, C. Dekker, L. P. Kouwenhoven, and H. S. J. van der Zant, Nature **429**, 389 (2004).
[21] A. Makarovski, L. An, J. Liu, and G. Finkelstein, Phys. Rev. B **74**, 155431 (2006).
[22] S. Ilani, L. A. K. Donev, M. Kindermann, and P. L. McEuen, Nature Phys **2**, 687 (2006).
[23] See Supplemental Material for a discussion of growth and electrical characterization, theoretical calculation of chemical potential, theoretical values of effective length and Bohr radius, and "exchange only" theory at http://link.aps.org/supplemental/10.1103/PhysRevLett.123.197701. Supplemental Material includes Refs [33,34].

[24] T. Sharf, J. W. Kevek, and E. D. Minot, in *2011 11th IEEE International Conference on Nanotechnology* (2011), pp. 122–125.
[25] N. Lotfizadeh, M. J. Senger, D. R. McCulley, E. D. Minot, and V. V. Deshpande, https://arxiv.org/abs/1808.01341 (2018).
[26] R. Berkovits and B. L. Altshuler, Phys. Rev. B **55**, 5297 (1997).
[27] F. Simmel, D. Abusch-Magder, D. A. Wharam, M. A. Kastner, and J. P. Kotthaus, Phys. Rev. B **59**, R10441 (1999).
[28] M. Bockrath, D. H. Cobden, P. L. McEuen, N. G. Chopra, A. Zettl, A. Thess, and R. E. Smalley, Science **275**, 1922 (1997).
[29] H. Ajiki and T. Ando, Journal of the Physical Society of Japan **62**, 1255 (1993).
[30] H. Ajiki and T. Ando, Journal of the Physical Society of Japan **65**, 505 (1996).
[31] E. D. Minot, Y. Yaish, V. Sazonova, and P. L. McEuen, Nature **428**, 536 (2004).
[32] M. M. Fogler, Phys. Rev. Lett. **94**, 056405 (2005).
[33] E. Pop, D. Mann, J. Cao, Q. Wang, K. Goodson, and H. Dai, Phys. Rev. Lett. **95**, 155505 (2005).
[34] L. Calmels and A. Gold, Phys. Rev. B **52**, 10841 (1995).

**Supplemental Material For "Band-Gap-Dependent Electronic Compressibility of Carbon Nanotubes in the Wigner Crystal Regime"**

Neda Lotfizadeh[1], Daniel R. McCulley[2], Mitchell J. Senger[2], Han Fu[3], Ethan D. Minot[2], Brian Skinner[4,5] and Vikram V. Deshpande[1]*

[1]Department of Physics and Astronomy, University of Utah, Salt Lake City, Utah 84112, USA.

[2]Department of Physics, Oregon State University, Corvallis, OR 97331, USA.

[3]James Franck Institute, University of Chicago, Chicago, Illinois 60637, USA.

[4]Department of Physics, Massachusetts Institute of Technology, Cambridge, MA 02139, USA.

[5]Department of Physics, Ohio State University, Columbus, Ohio 43210, USA.

*Correspondence to: vdesh@physics.utah.edu..

**Growth and Electrical Characterization:**

The iron catalyst pads (consisted of a 1-nm layer of Fe, a 20-nm layer of $SiO_2$ and 1-nm layer of Ti) were deposited about 1 μm from the edge of the Pt electrode. A fast-heat CVD process was used to preserve the integrity of the Pt electrodes. The chip is placed in a furnace and the process is carried out at 800 °C in hydrogen for 1 minute followed by a 2:1 mixture of methanol and ethanol vapor flowed over the chip for 5 minutes. After growth, about 10% of the electrodes are connected electrically by a single-walled carbon nanotube.

We test each nanotube by invoking the high-bias maximum current test of Eric Pop and coworkers [1]. According to their study, the maximum current in individual suspended carbon nanotube is approximately ~10μA/$L$, where $L$ is the length of the nanotube in microns, with some variation depending on nanotube diameter. We follow this trend to selectively choose individual nanotubes for our study and disregard multiple or multiwall nanotubes. Fig. S1 shows high-bias I-

V characteristic of large bandgap nanotubes (CN2-CN6). It can be seen that these tubes show the same trend and magnitude as ones in previous studies [1].

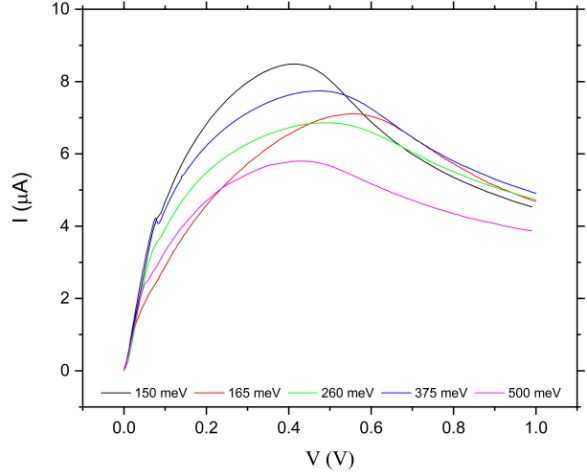

**Fig. S1: High-bias** I-V curves of CN2-CN6.

**Theoretical Calculation of the Chemical Potential:**

Within our Wigner crystal approximation, the electrostatic contribution to the chemical potential, $\mu_{el}$, is a function of only the electron concentration $n$ and the distance $d$ to the gate electrode:

$$\mu_{el} = \left(\frac{1}{L_{eff}}\right)\frac{dE_{el}}{dn} = \frac{2e^2 n}{4\pi\varepsilon_0}\sum_{i=1}^{\infty}\left(\frac{1}{i} - \frac{i^2 + 2(nd)^2}{[i^2 + (2nd)^2]^{\frac{3}{2}}}\right). \quad (1)$$

The kinetic energy of an electron in a box of width 1/$n$ is

$$E_K = N\left(\sqrt{(\pi\hbar vn)^2 + \left(\frac{\Delta}{2}\right)^2} - \frac{\Delta}{2}\right). \quad (2)$$

From this definition of $E_K$ one can define a kinetic energy contribution to the chemical potential

$$\mu_k = \frac{dE_K}{dN} = \frac{2(\pi\hbar v n)^2 + (\Delta/2)^2}{\sqrt{(\pi\hbar v n)^2 + (\Delta/2)^2}} - \frac{\Delta}{2}. \quad (3)$$

At scales that are long compared to $d$, one can consider the density and chemical potential as smoothly varying functions. In this case, the distribution $n(x)$ of electron density is that which makes the electrochemical potential spatially uniform

$$\mu_0 = \mu_{el}[n(x)] + \mu_k[n(x)] + U(x) = const. \quad (4)$$

In order to solve for $\kappa^{-1}$ as a function of electron number $N$, we fix the value of the electrochemical potential $\mu_0$ and solve for the density $n$ at each point $x$ by finding the root of Eq. (4) numerically, with the functions $\mu_{el}(n)$ and $\mu_k(n)$ given by Eqs. (1) and (3), respectively. The total number of electrons corresponding to the electrochemical potential $\mu_0$ is then $N = \int n(x)dx$. This procedure gives a relation between $\mu_0$ and $N$, resulting in the inverse compressibility defined by $\kappa^{-1} = d\mu_0/dN$.

Fig. S2 compares our experimental data (a) with theoretical results (b) for inverse compressibility of CN1 as a function of carrier number at different magnetic fields. It can be seen that our theory shows similar behavior as our experimental results but cannot achieve the same reduction in low densities at higher magnetic fields.

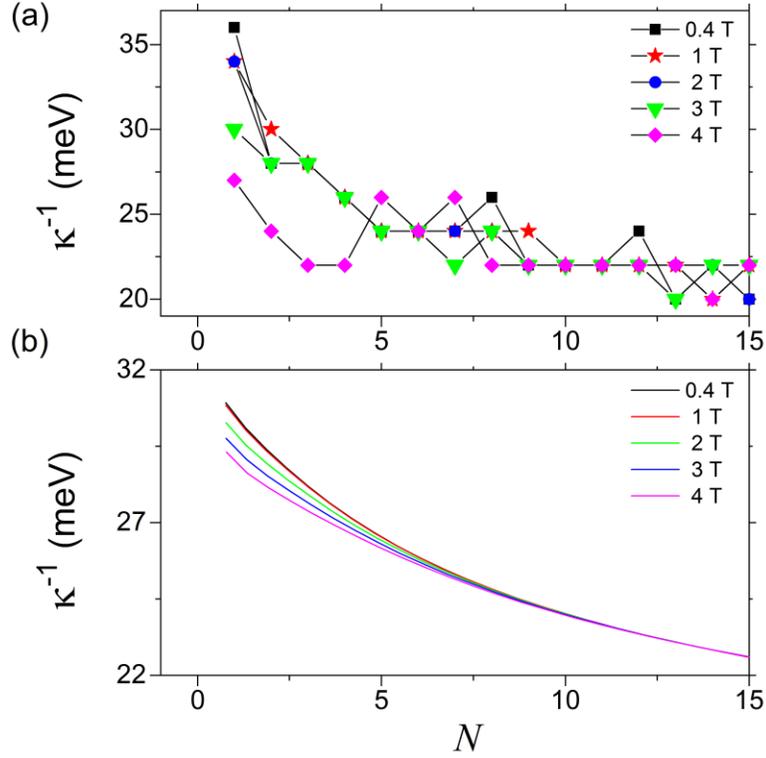

**Fig. S2:** Experimental data (a) and theoretical results (b) of $\kappa^{-1}$ as a function of carrier number ($N$) for CN1 at different magnetic fields. Similar changes in both results can be seen in $\kappa^{-1}$ as a function of magnetic field.

**Theoretical Values of Effective Length and Bohr Radius:**

Obtained values of $L_{\text{eff}}$ for individual tubes (CN1-CN6) are presented in Table S1. Since $L_{\text{eff}}$ is density dependent in the presence of a confining potential, we do not list $L_{\text{eff}}$ for CN1. $L_{\text{eff}}$ is smaller in tubes with larger bandgaps. The interaction strength can be estimated from average carrier spacing divided by effective Bohr radius, $r_s \approx 1/(na_B)$. Calculated values of $a_B$ from our theory are included in Table S1.

| Δ (meV) | 25 | 150 | 165 | 260 | 375 | 500 |
|---|---|---|---|---|---|---|
| $L_{eff}$ (nm) | - | 1232 | 1151 | 1085 | 932 | 681 |
| $a_B$ (nm) | 15 | 2.56 | 2.32 | 1.47 | 1.03 | 0.77 |

**Table S1.** Calculated effective length $L_{eff}$ and Bohr radius $a_B$ for tubes with different bandgaps. In tube with Δ= 25 meV, $L_{eff}$ depends on the electron density.

**Exchange-Only Theory:**

An alternative theoretical approach to the one considered in the main text is to consider the limit of weak interactions, $r_s = 0$, in which the system is described as a uniform electron gas with exchange interactions. The total energy of the 1D electron gas is given as the total kinetic energy of electrons $E_K$ plus the total Coulomb energy $E_C$ of the system. Even though the electron density is spatially uniform, the exchange interaction still guarantees a nontrivial correction to $E_C$ relative to the usual capacitance of a metallic wire. In the limit where the distance between electrons (L/N), is much larger than the distance $d$ to the gate electrode, this correction becomes large and can reduce the value of $\kappa^{-1}$.

The Fermi energy and the total kinetic energy (at $N \gg 1$) of a system with spin degeneracy of 2 is:

$$E_F = \frac{\hbar^2}{2m}\left(\frac{\pi N}{2L}\right)^2 \to E_K = \int_0^N E_F(N')dN' = \frac{\pi^2 \hbar^2 N^3}{24mL^2}.$$

To estimate the Coulomb energy, one can consider an electron at the origin and calculate its interaction energy $U$ with all other electrons, $U = 1/2 \int dr \rho(r) V(r)$, where $\rho(r)$ is the change density and $V(r)$ is the Coulomb interaction. The factor 1/2 in front of the integral prevents double-counting. In our problem, the relevant charge density $\rho(r)$ is described by the pair distribution function (PDF) $g(r)$. In particular, $g(r)$ is defined so that with an electron at the origin, the probability of finding another electron in the interval $(r, r + dr)$ is $eng(r)dr$, where $n = N/L$. Therefore, the total Coulomb energy is $E_C = (N^2/2L) \int V(r)g(r)dr$, where $V(r)$ is the gate-screened Coulomb interaction.

$$V(r) = \frac{e^2}{4\pi\varepsilon_0}\left(\frac{1}{r} - \frac{1}{\sqrt{r^2 + (2d)^2}}\right)$$

The integral of $E_C$ should be cut of at very small $x$, comparable to the wire radius $r_0$, at which the system ceases to be one-dimensional in an electrostatic sense. In a uniform Fermi gas, electrons with opposite spins pass freely through each other and effectively have $g(r) \equiv 1$. But electrons with the same spin cannot occupy the same position, and must have $g(r \to 0) = 0$. The combined PDF for both spins is described by $g(r) = 1 - \frac{1}{2}(\sin(k_F r)/k_F r)^2$ [2], where $k_F = \pi N/2L$. Taking second derivate of total energy $(E_C + E_K)$ yields:

$$\kappa^{-1} = \frac{e^2}{4\pi\varepsilon_0}[\frac{\pi^2 N a_B}{4L} + \int_{r_0}^{\infty}\left(\frac{1}{x} - \frac{1}{\sqrt{x^2 + (2d)^2}}\right)\left(2 - \cos\left(\frac{\pi N x}{2L}\right)\right)dx]$$

Fig. S3 compares the result of using exchange-only theory (a) and the measured data (b) for CN2-CN6. It can be seen that this theory can achieve the suppression in $\kappa^{-1}$ at low densities of large bandgap CNTs, which arises from a reduction in electrostatic energy associated with the exchange interaction. However, this model does not reproduce the Wigner crystal and we do not consider it appropriate for the large-$r_s$ limit that is the domain of our experiments.

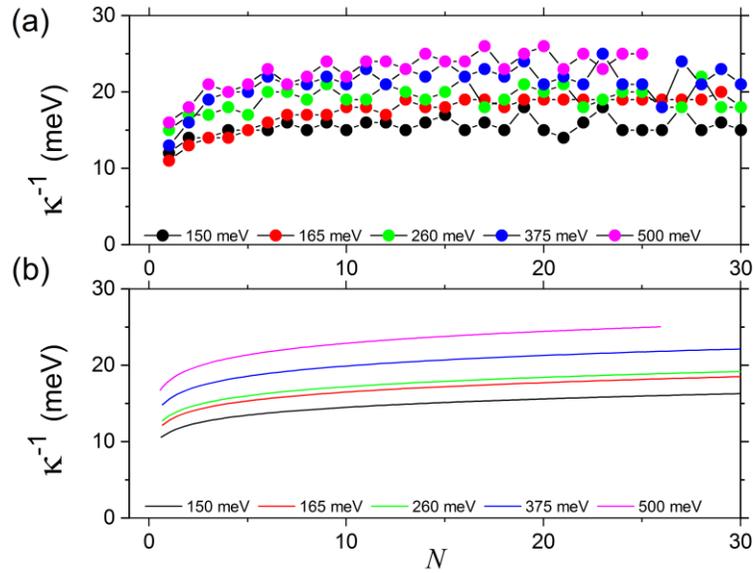

**Fig. S3:** Experimental data (a) and theoretical results (b) of inverse compressibility as a function of carrier number ($N$) for nanotubes with different bandgaps (CN2-CN6) using total energy of uniform 1D electron gas.

**References:**

[1] E. Pop, D. Mann, J. Cao, Q. Wang, K. Goodson, and H. Dai, Phys. Rev. Lett. **95**, 155505 (2005).
[2] L. Calmels and A. Gold, Phys. Rev. B **52**, 10841 (1995).